\documentstyle[aps,multicol,epsf]{revtex}
\draft
\newcommand{\beq}{\begin{equation}}
\newcommand{\eeq}{\end{equation}}
\newcommand{\bdis}{\begin{displaymath}}
\newcommand{\edis}{\end{displaymath}}
\newcommand{\bea}{\begin{eqnarray}}
\newcommand{\eea}{\end{eqnarray}}
\newcommand{\barr}{\begin{array}}
\newcommand{\earr}{\end{array}}

\begin{document}

\title{First-Order Transition
in the Breakdown of Disordered Media}

\author{Stefano Zapperi$^1$, Purusattam Ray$^2$,
H. Eugene Stanley$^1$ and Alessandro Vespignani$^3$}

\address{$^1$Center for Polymer Studies and Department of Physics,
        Boston University, Boston, Massachusetts 02215\\
        $^2$The Institute of Mathematical Sciences,
        CIT Campus, Madras - 600 113, India\\
        $^3$Instituut-Lorentz, University of Leiden, P.O. Box 9506, 2300 RA,
        Leiden, The Netherlands.}

\date{\today}

\maketitle

\begin{abstract}

We study the approach to global breakdown in disordered media driven
by increasing external forces.  We first analyze the problem by mean-field
theory, showing that the failure process can be described as a
{\em first-order} phase transition, similarly to the case of thermally
activated fracture in homogeneous media. Then we 
{\em quantitatively} confirm the predictions
of the mean-field theory using numerical simulations of
discrete models.  Widely distributed avalanches 
and the corresponding mean-field scaling are explained 
by the long-range nature of elastic interactions.  
We discuss the analogy of our results to driven disordered
first-order transitions and spinodal nucleation in magnetic
systems.

\end{abstract}

\date{\today}



\begin{multicols}{2}

The breakdown of solids under external forces is a longstanding problem
that has both theoretical and practical relevance \cite{breakdown}.  The
first theoretical approach to fracture mechanics, proposed by Griffith
\cite{griff} more than 75 years ago, is similar to the classical theory of
nucleation in first-order phase transitions \cite{nucl}.  An elastic
solid under stress is in a ``metastable state'' and will decay to the ``stable
fractured state'' by the formation of cracks. Recently
it has been shown \cite{buse} that the point of zero external stress has
the same mathematical properties as the condensation point in gas-liquid
first-order transitions. In the language of phase transitions, the stress
imposed on the solid plays the role of an external field and cracks are
the analog of droplets of a new phase. The classical theory of nucleation is
expected to fail close to the limit of stability, the spinodal point
\cite{sn} and it has been suggested \cite{rk,sel1,sel2,gol} that
a similar behavior should occur for fracture, for large values of
the external stress.  Thus, the failure threshold
corresponds to the spinodal point of first-order phase transitions.

The Griffith theory and related calculations deal with the situation in
which fracture is {\em thermally} activated and {\em quenched} disorder
is absent or weak.  In many realistic situations, however, the solid is
not homogeneous, and disorder, in the form of vacancies or microcracks,
strongly affects the nucleation process \cite{sel2,gol}.  There are
situations, encountered for example in material testing, in which the
system is driven by an increasing external stress \cite{jam,sorn3} and
the time scale of thermal fluctuations is much larger than the time
scale induced by the driving. In those cases, the system can be
effectively considered as being at zero temperature, so only quenched
disorder is relevant.  It has been experimentally observed
\cite{jam,sorn3,ae,ccc,dmp} that the response (acoustic emission) of
stressed disordered media takes place in bursts of widely distributed
intensity, indicative of an internal avalanche dynamics.

The understanding of the breakdown of disordered systems has 
considerably progressed due to the use of large scale simulations of lattice
models \cite{hr}. These models have provided a good description of
geometrical and topological properties of cracks, leading to the
introduction in this field of scaling concepts.  Recently, these models
have also been used to study the dynamic response of the system before
breakdown \cite{th,chak,spring,gcalda,sahimi}.  Scaling and power-law
avalanche distributions were observed, in agreement with experiments,
but a clear theoretical interpretation of the results is still lacking.
Here, we address the problem by mean-field calculations and
numerical simulations. The picture that emerges from our analysis is
that the breakdown in disordered media can be described by a first-order
transition, similarly to the case of thermally-activated homogeneous
fracture.  Since elastic interactions are long-range, scaling behavior
may be present also in low dimensions, in analogy with spinodal
nucleation \cite{sn}.

The models we will consider are defined for a two-dimensional lattice of
linear size $L$.  Each bond of the lattice represent an elastic spring
that breaks when it is stretched beyond a threshold chosen
from a given probability distribution. An
external stress is imposed on the system by suitably chosen boundary
conditions.  A simple example, because of the scalar nature of the
interactions, is the random fuse model \cite{fuse} for electric breakdown. 
To each bond $i$ is associated a resistor of unit
conductivity ($\sigma_i=1$).  When the current in the bond exceeds a
threshold $D_i$ the bond becomes an insulator ($\sigma_i=0$).  A slowly
increasing external current \cite{str-cur} is imposed on the lattice and
the voltage drops $(\Delta V)_i$ for each bond are computed by
minimizing the total dissipated energy
\beq
E(\{\sigma\})\equiv\frac{1}{2}\sum_i \sigma_i ((\Delta V)_i^2-D^{2}_i).
\label{en}
\eeq
The dynamics of the model results from a double minimization
process. The voltage drops $(\Delta V)_i$ are obtained by a {\em global}
minimization of the energy at fixed $\sigma_i$, while the $\sigma_i$ are
then chosen to minimize the {\em local} bond energy. This last step
corresponds to breaking the bonds for which the current overcomes the
threshold.  The external current is increased slowly until the lattice
is no longer conducting. We note that this dynamics is very similar in
spirit to that of a field-driven random-field Ising model (RFIM),
studied by Sethna et al. \cite{set1,dahm} in the context of magnetic
hysteresis. In that model, each spin chooses the sign of the local field.

To derive a mean-field theory, it is useful to recast the dynamics of
the model in terms of the externally applied current $I$. In full
generality, we can rewrite the energy of Eq.~(\ref{en}) as
\beq
E(I,\{\sigma\})=\frac{1}{2}\left(\frac{I^2}{G(\{\sigma\})},
-\sum_i\sigma_i D^{2}_i\right)
\eeq
where $G(\{\sigma\})$ is the total conductivity of the lattice and is a
complicated function of the local conductivities.  We can estimate 
$G(\{\sigma\})$ using the effective medium theory \cite{kirk}, which
in our case gives
\beq
G(\{\sigma\})=\phi-A\phi(1-\phi)+O((1-\phi)^2),
\label{eq:effmd}
\eeq
where $\phi\equiv\sum_i \sigma_i/L^2$, and $A\simeq 1.52$.  Using only
the first term of the expansion, we can express the energy as a sum over
``spins'' interacting with effective random fields $h_i$
\beq
E_{MF}(I,\{\sigma\})=\sum_i\sigma_i h_i=\frac{1}{2}\sum_i \sigma_i
\left(\frac{I^2}{(L\phi)^2}-D_i^2\right).
\eeq
the value of $\phi$ can be computed self-consistently as
\beq
\phi = P(h_i>0)=1-\int_0^{\frac{I}{L\phi}}\rho(D)dD, 
\label{mfphi}
\eeq
where $\rho(D)$ is the distribution of failure thresholds. 
The solution of this equation can be expressed in terms
of the current per unit length $f\equiv I/L$.
We can identify $f$ with the external field and $\phi$
with the order parameter. 

{}From considerations similar to those presented in Ref.~\cite{dahm}, we can
show that for any analytic \cite{square} normalizable distribution
function $\rho$ Eq.~(\ref{mfphi}) has a solution for $f<f_c$ and, close to
$f_c$, $\phi$ scales as
\beq
\phi - \phi_c \sim (f_c-f)^{1/2}.
\label{scalphi}
\eeq
This scaling law is also valid if we include higher order terms from
Eq.~(\ref{eq:effmd}).  The mean-field theory we have presented can be
exactly mapped to the democratic fiber-bundle model (DFBM), an exactly
solvable model for fracture which has been studied extensively 
\cite{dfbm}.  We can, therefore, obtain the mean-field avalanche
size distribution from the exact results derived for the DFBM
\cite{dfbm}
\beq
P(m)\sim m^{-\tau}f(m(f_c-f)^\kappa); ~~~~~\tau=\frac{3}{2},~~\kappa = 1,
\label{eq:pm}
\eeq
where $m$ is the number of bonds that break as function of the current.
The average avalanche size $\langle m \rangle$ is proportional to
the ``susceptibility'' $d\phi/df$ \cite{chak}, and therefore diverges at the
breakdown as
\beq
\langle m \rangle \sim (f_c-f)^{-\gamma}~~~~~~\gamma=1/2.
\label{susc}
\eeq
The exponents we have introduced satisfy the scaling relation
$\kappa(2-\tau)=\gamma$, which is consistent with the values reported in
Eq.~(\ref{eq:pm}) and Eq.~(\ref{susc}).
The mean-field analysis indicates that the system is undergoing a first
order transition since the order parameter has a discontinuity and
the conductivity at $f_c$ has a {\em finite\/}
jump from $G(\phi_c) > 0$ to zero. The approach to this transition is
characterized by avalanches of increasing size, diverging at the
transition.  

A similar behavior with the same scaling
exponents is observed in the mean-field theory of the driven RFIM
\cite{set1,dahm} for small disorder. In the RFIM, one observes also a
second order transition as the width of the disorder is increased. A
similar transition does not seem to be present in our system, at least
not in the mean-field treatment.  
It is also interesting to note that the same
scaling laws describe metastable systems close to a spinodal point.  The
quasistatic susceptibility diverges as in Eq.~(\ref{susc}) and droplets
are distributed according to Eq.~(\ref{eq:pm}).

\begin{figure}[htb]
\narrowtext
\centerline{
        \epsfxsize=7.0cm
        \epsfbox{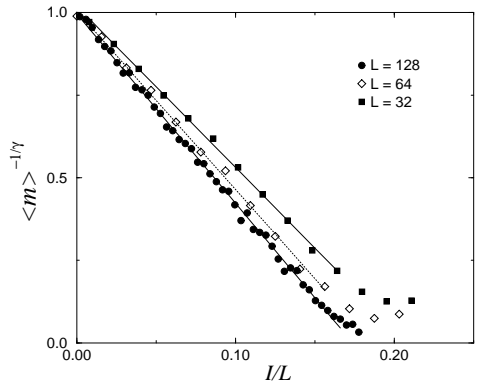}
        \vspace*{0.5cm}
        }
\caption{The average avalanche size $\langle m \rangle^{-1/\gamma}$
is plotted as a function of $I/L$, using the mean-field value 
$\gamma=1/2$. The curves are averaged over several realizations
of the disorder ($N=500$ for $L=32$, $N=100$ for $L=64,128$).
The critical current of the random fuse model $I_c/L$ has logarithmic
corrections, so the curves do not superpose.}
\label{fig:1}
\end{figure}

\begin{figure}[htb]
\narrowtext
\centerline{
        \epsfxsize=7.0cm
        \epsfbox{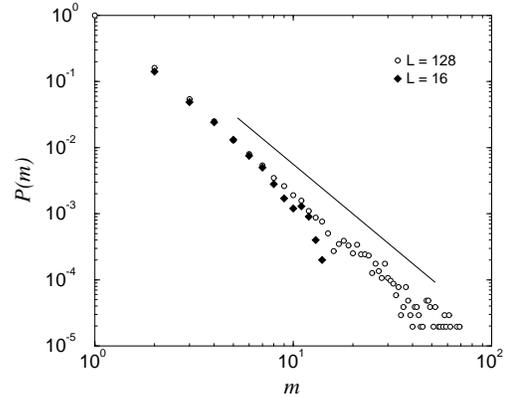}
        \vspace*{0.5cm}
        }
\caption{The avalanche size distribution for a random fuse model
of size $L=32$ and $L=128$ is plotted in log-log scale. 
A line with the mean field value $\tau^{\prime}=5/2$ 
of the exponent is plotted for reference.
The cutoff of the distribution increases with the system size.}
\label{fig:2}
\end{figure}

An important issue to address at this point is the validity of
mean-field results in the case of real low-dimensional systems.
It is known that scaling does not hold close to the first-order transition
for short-ranged RFIM in dimensions $d=2,3$ \cite{set1,dahm}.
Similarly, spinodal singularities are observed when interactions are
long-range \cite{raykl}.  Elastic interactions are
intrinsically long-range, which leads to mean-field behavior even for low
dimensions, as we will next show numerically.
We simulate the random fuse model \cite{fuse} on a tilted square
lattice, with periodic boundary conditions in the transverse
direction. The current in each bond is computed numerically by solving the
Kirchhoff equations with a precision $\epsilon = 10^ {-10}$. The
distribution of thresholds is chosen to be uniform in the interval
$[1-\Delta,1+\Delta]$. We made the choice $\Delta=1$ in order to avoid
the ``ductile-brittle'' crossover at a finite value of the lattice size
\cite{fuse}. Other broad analytic distributions give rise to similar
results.  We impose an external current $I$ through the lattice and we
increase it at an infinitesimal rate. When a bond fails, we recompute
the currents to see if other failures occur.  The system responds to the
increase of the current with avalanches whose size $m$ diverges at the
breakdown $f_c$.  In Fig.~\ref{fig:1} we plot the average avalanche size
$\langle m \rangle$ versus $I/L$ and we see that the mean-field exponent
$\gamma =1/2$ fits the data quite well. The data collapse is not perfect
because logarithmic corrections are expected in $d=2$ \cite{fuse}.  The
avalanche size distribution, {\em integrated} over the entire range of
the current, is plotted in Fig.~\ref{fig:2}.  We see that the scaling is
consistent with an exponent $\tau'=\tau+1/\kappa=5/2$, which results
from the mean-field calculations [Eq.~(\ref{eq:pm})].  As expected the
cutoff of the distribution increases with $L$.

We have studied the behavior of the average crack size $\langle s
\rangle$ as a function of the current.  Although the crack size grows
with the current, it does not appear to diverge at the breakdown when
the system size is increased. This implies that the final macroscopic
crack is formed by the {\em coalescence} of several microcracks, rather
than by the growth of a single crack. In fact, by monitoring the dynamics,
we observe that avalanches are not spatially
connected since interactions are long-range; this may be the reason 
why mean-field works so well for these systems.  Finally, we
have checked that the conductivity of the lattice displays a finite jump
at the first-order transition.

\begin{figure}[htb]
\narrowtext
\centerline{
        \epsfxsize=7.0cm
        \epsfbox{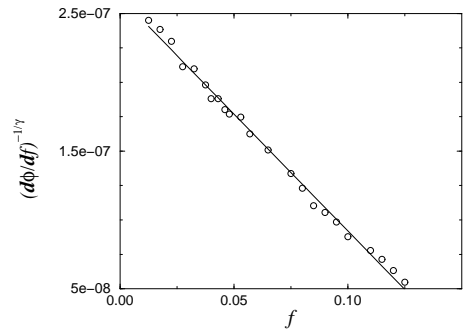}
        \vspace*{0.5cm}
        }
\caption{The suscptibility $d\phi/df$ for the  spring network
of size $L=50$, averaged over $N=100$ configurations,
is plotted as function of the 
applied stress $f$. Mean-field scaling ($\gamma=1/2$) 
appears to be very good also in this case.}
\label{fig:3}
\end{figure}

The mean-field theory was derived for the case of a scalar model,
but the results do not only apply to scalar models.
We have numerically simulated a more complex vectorial model, defined in
Ref.~\cite{spring}. The model is a spring network with central and bond
bending forces, with random failure thresholds associated with each
bond. The system is driven by an increasing external stress $f$ and the
dynamics is obtained by numerically integrating the equations of motion
of the springs.  The model gives a more reliable description of the
fracture process, taking into account the tensorial nature of the
elastic interactions and a realistic relaxation dynamics. In this
case we also find power-law distributed avalanches and mean-field exponents
(see Fig.~\ref{fig:3}).  A detailed account of the results of this model,
as well as a complete discussion of the random fuse model, will be
reported elsewhere \cite{long}.

The results we have discussed clarify the nature of the breakdown
process in the presence of quenched disorder.  We have shown that the
breakdown corresponds to a first-order transition, with avalanche
precursors characterized by power laws. The scaling exponents are in
quantitative agreement with the prediction of mean-field calculations.
In mean-field theory, driven disordered systems behave similarly to their
homogeneous, thermally driven, counterparts, if we compare the scaling
of avalanches with that of the droplets.  This applies to the RFIM
\cite{set1,dahm}, which shows features similar to those of spinodal
nucleation \cite{sn}, and to the fracture models we have
studied. However, one should be careful not to interpret these analogies
too strictly, since in driven disordered systems the notions of
metastability, spinodal point and nucleation are not well defined.

Finally, we comment on the applicability of 
self-organized criticality (SOC) \cite{soc} to
fracture problems.  In fact,  first-order transitions and
SOC are becoming the principal competing theoretical frameworks for
the interpretation of avalanche phenomena in disordered systems.  A
striking example of this controversy is represented by earthquake
phenomena \cite{earth}. The definition of SOC
implies a slowly driven system with a critical {\em stationary}
state\cite{vzl}. The only possibility to
observe SOC behavior in fracture phenomena is in the presence of a long
lived plastic state before rupture.  Recently proposed scalar models
\cite{zvs} of microfractures and molecular dynamics simulations of
granular solids under shear \cite{gran} have shown that
the plastic stationary state is characterized by power law
avalanche distributions suggestive of SOC. 
On the other hand, the model studied in Ref.~\cite{gcalda},
that was claimed to display SOC, 
has no stationary state, like the models discussed
here. Power laws without cutoff in this case arise not due to
self-organization, but because the control parameter is externally
``swept'' towards the instability (as pointed 
out by Sornette \cite{sornette}). 
We believe that different experimental
conditions can all give rise to similar scaling behavior, 
but the underlying physical mechanisms could be quite different. 

The Center for Polymer Studies is supported by the NSF.  We thank
P. Cizeau, W. Klein and F. Sciortino for interesting discussions and
remarks.

\end{multicols}

\begin{thebibliography}{99}
\bibitem{square}
        The results are also valid in the case of the 
        uniform distribution we use in simulations, 
        as can be shown by a direct calculation.
\end{thebibliography}
\end{document}